\documentclass[3p,sort&compress,number]{elsarticle}

\usepackage{graphicx}
\usepackage{lineno}
\usepackage[hidelinks]{hyperref}
\usepackage{setspace}
\modulolinenumbers[5]

\journal{Acta Materialia}

\newcommand{\degree}{^{\circ}}
\begin{document}
\doublespacing

\begin{frontmatter}
\title{Nucleation and growth of hierarchical martensite in epitaxial shape memory films}

\author[1]{R. Niemann}

\author[1]{A. Backen}
\author[1]{S. Kauffmann-Weiss\corref{a}}
\cortext[a]{current address: Karlsruher Institute of Technology, Institute for Technical Physics, 76344 Eggenstein-Leopoldshafen, Germany}

\author[1]{C. Behler}
\author[1]{U. K. R\"o\ss ler}
\author[2]{H. Seiner}
\author[3]{O. Heczko}
\author[1,4]{K. Nielsch}
\author[1,4]{L. Schultz}
\author[1]{S. F\"ahler\corref{mycorrespondingauthor}}
\cortext[mycorrespondingauthor]{Corresponding author}
\ead{s.faehler@ifw-dresden.de}

\address[1]{IFW Dresden, Helmholtzstra\"sse 20, 01069 Dresden, Germany}
\address[2]{Institute of Physics, Czech Academy of Sciences, Na Slovance 2,
18221 Prague, Czech Republic, Na Slovance 2, 18221 Prague, Czech Republic}
\address[3]{Institute of Thermomechanics, Czech Academy of Sciences, Dolejskova 5, 18200 Prague, Czech Republic}
\address[4]{Technische Universit\"at Dresden, Institute of Materials Science, 01062 Dresden, Germany}

\hyphenation{theo-retical}

\begin{abstract}
Shape memory alloys often show a complex hierarchical morphology in the martensitic state. To understand the formation of this twin-within-twins microstructure, we examine epitaxial Ni-Mn-Ga films as a model system. In-situ scanning electron microscopy experiments show beautiful complex twinning patterns with a number of different mesoscopic twin boundaries and macroscopic twin boundaries between already twinned regions. 
We explain the appearance and geometry of these patterns by constructing an internally twinned martensitic nucleus, which can take the shape of a diamond or a parallelogram, within the basic phenomenological theory of martensite. These nucleus contains already the seeds of different possible mesoscopic twin boundaries. Nucleation and growth of these nuclei determines the creation of the hierarchical space-filling martensitic microstructure.
This is in contrast to previous approaches to explain a hierarchical martensitic microstructure. This new picture of creation and anisotropic, well-oriented growth of twinned martensitic nuclei explains the morphology and exact geometrical features of our experimentally observed twins-within-twins microstructure on the meso- and macroscopic scale.
\end{abstract}

\begin{keyword}
shape memory\sep martensite \sep nucleation \sep Ni-Mn-Ga
\end{keyword}

\end{frontmatter}


\section{Introduction}

\begin{figure*}[tb]
	\centering
		\includegraphics[width=\textwidth]{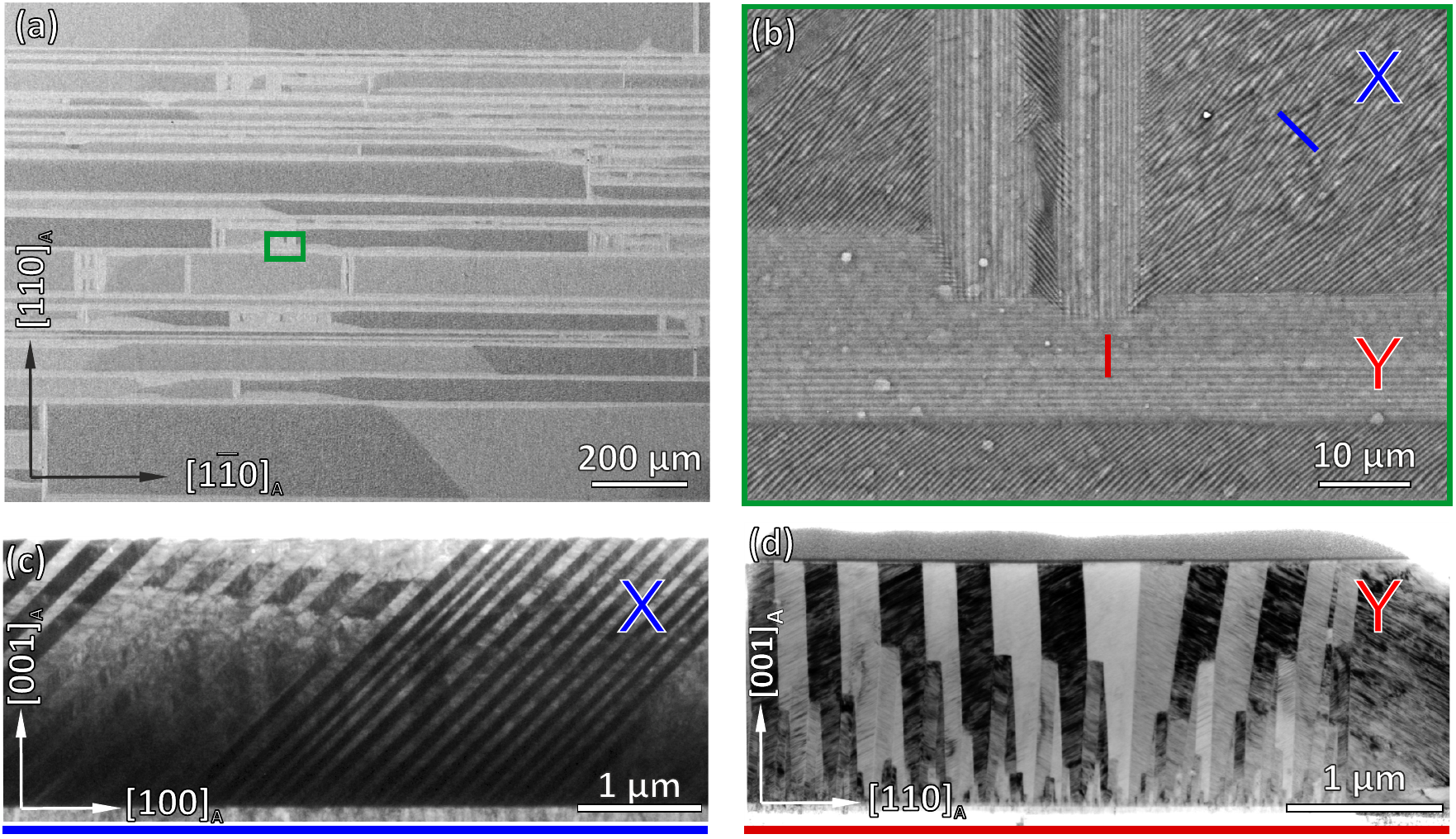}
	\caption{Hierarchical twin-within-twins martensitic microstructure. (a) SEM micrograph (backscattered electron contrast) of an epitaxial Ni-Mn-Ga film in the martensitic state at room temperature. The surface has a complex yet regular microstructure, dominated by macrotwin boundaries. (b) A magnification  at the area marked in green in (a) shows the characteristic of two microstructures called type X and type Y. All contrast stems from mesoscopic twin boundaries due to topography (type X) and variant orientation (both types).(c) is a TEM cross-section of the film in the region marked by the blue line in (b) of type X, (d) of type Y marked by the red line. 
	}
	\label{fig:Fig1}
\end{figure*}

Martensitic microstructures often exhibit a hierarchical twin-within-twins microstructure that displays a corresponding hierarchy of typical lengths, which often span several orders of magnitude. While such hierarchy can exhibit  beautiful patterns (e.g. Fig. \ref{fig:Fig1}), this microstructure and the different types of twin boundaries are critical for several functional properties of martensitic materials. One example is provided by Ni-Mn-Ga magnetic shape memory alloys \cite{Ullakko_APL96} where the presence of either type I or type II twin boundaries within the martensitic microstructure results in twinning stresses \cite{Straka2011b} and mobilities differing \cite{Faran2013} by an order of magnitude. Another key functional property depending on the microstructure is the transformation hysteresis loss, which must be as low as possible for magnetocaloric and elastocaloric refrigeration \cite{Faehler2012}.

To understand the formation of this hierarchical microstructure, we used epitaxial Ni-Mn-Ga films as a model system. All types of Ni-Mn-X (X = Ga, In, Sn, Al) films show a characteristic hierarchical microstructure \cite{Kaufmann2011, Leicht2011, Luo2011, Erkartal2012, Yang2014a, Auge2012, Yang2013,Teichert2015} and are of scientific and technological interest due to their magnetic shape memory \cite{Backen2013a, Niemann2010a, Ranzieri2015}, magnetocaloric \cite{Teichert2015, Auge2012, Ranzieri2013, Diestel2015}  and multicaloric \cite{Schleicher2015} properties. The absence of large angle grain boundaries makes epitaxial films similar to single crystals, but easier to grow and with a well-defined and clean surface right after deposition. The high surface-to-volume ratio of a film allows taking examinations of the surface as representative for most of the sample. Moreover, the rigid boundary condition to the substrate is beneficial for analysis since it provides a fixed reference frame. The high aspect ratio of a film gives a good statistical representation of nucleation during a martensitic transformation as the thick, rigid substrate elastically decouples most of the thin film from itself. 

In this work, we will  demonstrate that the shape of the nucleus is decisive for the formation of a hierarchical microstructure. For this we first describe briefly the final microstructure. Then we introduce two related geometrical shapes of nuclei: A diamond and a parallelogram. Finally, we present in-situ experiments, which illustrate how these nuclei form the hierarchical microstructure. Before coming to the results, we summarize the theoretical concepts and the state of art so far.

\subsection{Phenomenological theory of martensite}
A common concept to describe the microstructure of martensites is the phenomenological theory of martensite, which was developed by Wechsler, Lieberman and Read \cite{Wechsler1953}, and independently by Bowles and MacKenzie \cite{Bowles1954}. In this continuum description of crystallography it is assumed that all stress occurring during a symmetry-breaking phase transition is concentrated at the phase interface, and is minimized by introducing twin boundaries in the martensite. An accurate description of the microstructure after the phase transition is often  achieved by postulating that all twin boundaries and phase interfaces have to be invariant planes under the transformation. Modern descriptions of the formalism were published e.g. by Bhattacharya \cite{Bhat2003} or Pitteri and Zanzotto \cite{Pitteri2003}. The boundary conditions of invariant interfaces can be written in a matrix formalism. As input parameters, only the symmetry and the lattice constants of both phases are required. The geometric constraint at the boundary conditions yields the fraction of martensite twin variants along an invariant plane forming the phase boundary in addition to the relative orientation of both phases, all martensitic variants and all interfaces. The phase boundaries are planar and called habit planes. 
Due to its continuum character, the theory cannot provide the absolute size or spatial distribution of the microstructural features or information about the transformation path, but the absence of a scale makes it useful to describe twin-within-twin microstructures of higher order. 
Here we apply this theory to the transition from cubic austenite to a monoclinic 14M martensite as appropriate for the Ni-Mn-Ga alloy film used for our experiments.  For a descriptive understanding of the complete paper it is sufficient to keep in mind we consider that habit planes are formed by a combination of nanotwinning and related $a$-$b$-twinning in the modulated phase (see \cite{Gruner2016} for a recent discussion). The calculation of the habit plane for this particular system is described in detail elsewhere \cite{Niemann2016a}. They are close to  $\{110\}_\mathrm{A}$-planes, but differ by a few degrees.

\subsection{Modulation and adaptive nanotwinning}
The modulations of martensite can be considered a part of the hierarchical martensitic microstructure: According to the classical concepts of martensite, the twinning periodicity is a result of total energy minimization, which contains the elastic (volume) energy and the twin boundary energy. Following the concept of adaptive martensite \cite{Khach1991a} in case of a huge elastic and low twin boundary energy the twin boundary periodicity can be reduced to the nanoscale in order to adapt at the phase boundary to the austenite. Accordingly modulated structures can be considered as nanotwinned. The adaptive concept was also successfully applied for the particular 14M modulation in Ni-Mn-Ga examined here \cite{Kaufmann_PRL10, Kaufmann2011, Niemann2012a}. Recently also the regular arrangement of nanotwins within a modulation and the formation of $a$-$b$-twin boundaries has been explained by an interaction energy between nanotwins \cite{Gruner2016}. Thus a modulated structure can be considered as the first generation of twinning within a hierarchical martensitic microstructure.

However, there are competing explanations of modulated martensite and throughout the rest of this paper we will use the 14M unit cell as starting point to analyze the appearance of the hierarchical martensitic microstructure at the meso- and macroscale. In other words, this manuscript is not directly discussing the shortest length scale of twinning at the nanoscale. The construction of the modulated phases by nanotwinning is discussed elsewhere \cite{Niemann2016a, Gruner2016, Zeleny2016, Kaufmann_PRL10}. 

\subsection{Hierarchical martensitic microstructures}
Although there are several experimental observations of twin-within-twin microstructures in various martensites \cite{Inamura2011, Xie1998, Roytburd_93, MURAKAMI1994, Bhat1992} and in particular in martensitic Heusler alloys \cite{Straka2011, Chulist2014, Kaufmann2011, Leicht2011, Luo2011, Erkartal2012, Yang2014a, Auge2012, Yang2013, Teichert2015, Backen2013, Heczko2017}, there are just a few theoretical concepts explaining the entire microstructure. Phenomenological theory of martensite allows to construct compatible twin-within-twin microstructures \cite{Bhat2003}. This theory does neither consider length scales nor energies, hence it cannot explain why that many different types of twin boundaries should be introduced, covering several orders of magnitude. Roytburd \cite{Roytburd_93} proposed an alternative explanation, which considers that e.g. twinning of tetragonal martensite compensates linear elastic stress, but also creates some shear stress by bending at the twin boundary. To compensate this higher order term of elastic energy, a next generation of twinning is required and so on. Thus the hierarchical microstructure should be a result of global minimization of energy. Accordingly, the microstructure should rearrange until each twin boundary found its position of minimum energy. Neither of these concepts considers that a martensitic first order phase transformation proceeds by nucleation and growth. 

\section{Experimental details}
Direct-current magnetron sputtering under optimized conditions was used for film growth (substrate temperature: $573\,\mathrm{K}$, sputter power: $100\,\mathrm{W}$, base pressure: $10^{-9}\,\mathrm{mbar}$ range, working pressure: $0.008\,\mathrm{mbar}$ Ar) [21]. Two films were grown. Using a stoichiometric $\mathrm{Ni_2MnGa}$ standard, the composition was determined to be $\mathrm{Ni_{51}Mn_{23}Ga_{26}}$ with an error below $0.5\,\mathrm{at.\%}$ from energy-dispersive X-ray measurements. The thickness of the first film was about $5\,\mathrm{\mu m}$, the second $1.5\,\mathrm{\mu m}$ as determined by transmission electron microscopy. 
Micrographs were acquired in a scanning electron microscope (SEM, Zeiss LEO 1530 Gemini) equipped with a backscatter electron detector. The temperature was varied in the SEM using a Kammrath \& Weiss heating stage with temperature control. 
X-ray reciprocal space mapping (not shown) was used to determine the crystal structure. Film 1 is in monoclinic 14M phase at room temperature with the lattice parameters $a=0.615\,\mathrm{nm}$, $b=0.579\,\mathrm{nm}$. $c=0.554\,\mathrm{nm}$ and $\gamma=90.4\degree$ (in $L2_1$-derived convention \cite{Niemann2016a}). Film 2 is also in the 14M phase with $a=0.616\,\mathrm{nm}$, $b=0.582\,\mathrm{nm}$. $c=0.547\,\mathrm{nm}$ and $\gamma=90.2\degree$. The austenite lattice constants measured just above martensite start temperature resulted to $a_\mathrm{A}=0.582\,\mathrm{nm}$ for film 1 and $a_\mathrm{A} =0.585\,\mathrm{nm}$ for film 2.

\section{Results}

\subsection{Martensitic microstructure of epitaxial Ni-Mn-Ga films}
Figure \ref{fig:Fig1}a shows a scanning electron micrograph of the typical surface morphology of a Ni-Mn-Ga film in the martensitic state at low magnification. Large rectangular areas are separated by thin stripes. We will call boundaries between these large regions ``macroscopic'' twin boundaries. In Fig. \ref{fig:Fig1}b, we show a magnification of the marked green rectangle from Fig. \ref{fig:Fig1}a. The rectangular areas and the thin stripes show a different morphology. At the film surface, we can distinguish two types of patterns. The first type of microstructure, marked with an ``X'' is called ``type X'' \cite{Backen2013a, Ranzieri2015} and is also known as ``high contrast zone'' because of its pronounced topography contrast \cite{Yang2013}. 
Type X consists of twin bands aligned by $45\degree$ to the picture edges, along the $[100]_\mathrm{A}$ and $[010]_\mathrm{A}$ directions. They are not exactly parallel but form groups enclosing an angle of about $12\degree$, which leads to the appearance of rhombus-like features. This microstructure exhibits a characteristic zig-zag surface morphology \cite{Buschbeck2009}.
In the TEM cross-section (Fig. \ref{fig:Fig1}c), there are parallel twin bands enclosing an angle of about $45\degree$ with the substrate normal. The orientation of twin boundaries in the type X pattern are approximately $(101)_\mathrm{A}$, $(\bar{1}01)_\mathrm{A}$, $(011)_\mathrm{A}$ and $(0\bar{1}1)_\mathrm{A}$ planes. Their deviations from these high-symmetry planes are explained later in the framework of type I and type II twins. These planes are called \emph{mesoscopic}, because it is a twin of higher order than the nanotwins and can be observed e.g. in the SEM. It is one of the twin boundaries in Ni-Mn-Ga that can be moved by magnetic fields \cite{Seiner2014}. A detailed discussion of the types and relative twin orientation of these mesoscopic twin boundaries in Ni-Mn-Ga was published by Straka et al. \cite{Straka2011b}.

The second type of microstructure is marked with an ``Y'' and called ``type Y'' \cite{Backen2013a, Ranzieri2015} or ``low contrast zone'', because of its almost flat topography \cite{Yang2013}. There, the bands on the surface are aligned exactly along $[110]_\mathrm{A}$ or $[1\bar{1}0]_\mathrm{A}$. In the TEM cross-section (Fig. \ref{fig:Fig1}d), the traces of the mesoscopic twin boundaries are oriented approximately along the $[001]$-direction. Near the substrate, a significant branching of twin boundaries is observed. The twin boundary orientation is thus approximately, but not exactly, along $(110)$ and ($1\bar{1}0$) planes. The orientation of the martensitic variants in type X and type Y was recently studied by electron backscatter diffraction on a sample from the same series \cite{Niemann2016}. The twins could be identified as $c$-$a$-twins of modulated 14M martensite, with the $b$-axis lying in the film plane for type X and pointing out of plane for type Y. While this explains the different orientations of mesoscopic twin boundaries with respect to the surface, it does neither explain the rhombus nor why branching only occurs in type Y, and, most important, how all these features form.

\subsection{Geometry of compatible nuclei and their growth stages}
\begin{figure*}[tb]
	\centering
		\includegraphics[width=\textwidth]{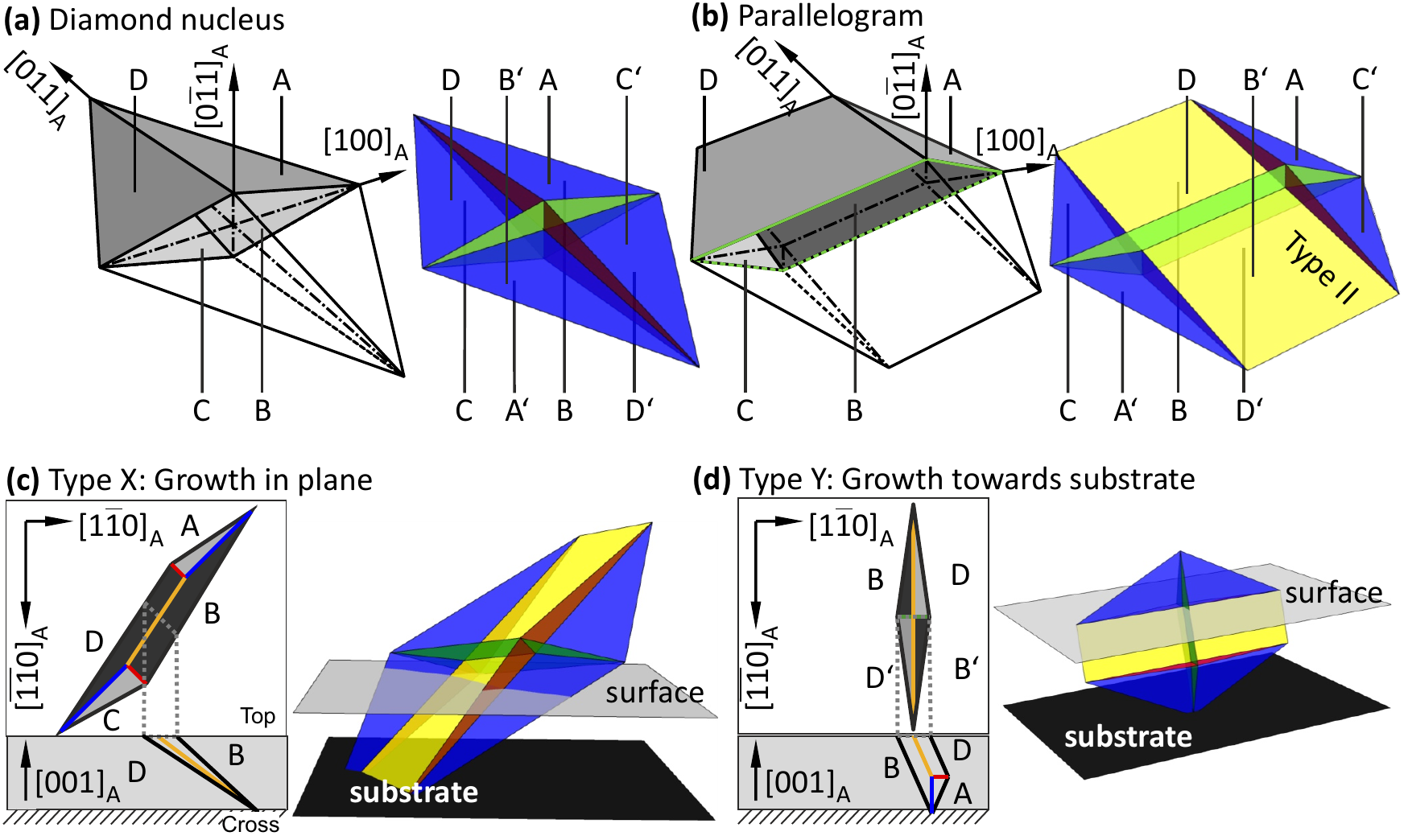}
	\caption{Model of the nucleus of martensite and its growth modes (a) A nucleus in the volume is enclosed by eight habit planes, all deviating a few degrees from the $(110)_\mathrm{A}$ plane. They are connected to each other by the cubic symmetry of the surrounding parent phase. Inside the nucleus, interfaces between the habit plane variants have to form which are mesoscopic twin boundaries, type I twin boundaries in blue and green and a modulation boundary in red. (b) The nucleus can grow in one direction while keeping all interface orientations constant. In this case, a new interface forms between B and B' and between D and D', respectively, which is a type II twin boundary (yellow). (c) The parallel growth mode in the setting of the type X-microstructure. The grey region represents a film cross-section, above the film surface is sketched. The right hand figure is a 3D view. The growth takes place in the film plane and the traces of the inclined type II twin boundary appear at the surface. (d) In type Y, the parallelogram grows perpendicular to the surface, hence the inclined type II twin boundary will be observed in the cross-section. For better clarity, the aspect ratio has been reduced in these images, hence all angles are not to scale.}
\label{fig:Fig2}
\end{figure*}

Martensitic transformations are diffusionless transformations of first order. Both phases have to coexist and thus form compatible phase boundaries. We will show in the following that the phenomenological theory of martensite predicts complex geometry for the nuclei, which differ significantly from spheres often considered in nucleation studies. Starting point for all geometries is the habit plane connecting austenite and martensite. The habit plane orientations can be calculated from the lattice parameters only \cite{Niemann2016a} and are close to $\left\{110\right\}$ planes \cite{Bhat2003}, but deviate by several degrees depending on the lattice constants and on the type of modulation. 

Due to the high symmetry of the parent phase, multiple equivalent solutions for the phase boundary are possible. These different solutions represent habit planes that are related to each other by the symmetry operations in the cubic phase, e.g. mirroring at $\{100\}_\mathrm{A}$ or $\{110\}_\mathrm{A}$ planes. 

Eight of these solutions, e.g. those close to the $(1\bar{1}0)$ plane, can enclose a volume of martensite (Fig. \ref{fig:Fig2}a). If all phase boundaries have the same extension, this leads to a martensitic volume in the shape of an elongated diamond (Fig. \ref{fig:Fig2}a). This is the most simple way to encapsulate martensite within austenite and thus we assume that this is the initial shape during formation of martensite. Its aspect ratio can be calculated from the known habit plane orientation and is in the order of 50 to 10 to 1. 

The eight habit planes connect the cubic austenite to eight different martensitic variants. These \emph{habit plane variants} are internally twinned as described in 1.2. When the internal twinning is disregarded, they can be described on the using a monoclinic 14M lattice. They are indexed in Fig. \ref{fig:Fig2}, with capital letters A, ..., D and A', ..., D'. Each pair A and A', B and B', ... shares the same crystal orientation, but the relative position of austenite and martensite at the habit plane is interchanged (their habit planes have antiparallel normals). All the variants have twin relationships to each other and some share twin boundaries, e.g. a mesocscopic type I $c$-$a$-twin boundary between A and D. Thus the need to form compatible boundaries between the different habit plane variants results in the formation of mesoscopic twin boundaries; the dark red plane in Fig. \ref{fig:Fig2}a and the blue plane are type I twin boundaries while the green plane is a modulation boundary. When cut by a surface, the traces of these twin planes become apparent as the midribs of the diamond. 

In a twinned nucleus, the faces that are formed by different habit planes can also have different sizes, which creates more complex and less symmetric shapes of the enclosed martensitic volume (Fig \ref{fig:Fig2}b). In this specific case, the face for the four habit plane variants B, B' and D, D' are larger than the other four. In the cross-section, this object looks like a long parallelogram. In the supplementary, we present an animation of the transformation from diamond to parallelogram.  We will later present an experimental proof that this is the growth stage of the original diamond nucleus. Diamond and parallelogram objects are crystallographically well-defined because the habit plane orientation does not change during growth in the parallelogram mode. 

Within the parallelogram, D and D' variants share a twin boundary, which they did not in the original nucleus, where they just met in a single point. This boundary is a $c_\mathrm{14M}$-$a_\mathrm{14M}$-boundary of type II and can be moved by even smaller forces than type I for bulk \cite{Straka2011b}. Also in thin films, magnetic-field induced reorientation was observed \cite{Thomas_NJP08}. With the knowledge of today, this effect can be ascribed to movement of type II twin boundaries of the type X microstructure. 

In Fig. \ref{fig:Fig2}c the orientation of a parallelogram is shown for type X microstructure in a film. In this sketch, the tip of the diamond is fixed at the interface between film and substrate. This is reasonable because the substrate-film interface is incompatible to martensite and thus the nucleus cannot grow further. At the surface, the parallelogram cross-section is obvious. 
In the setting of type Y (Fig. \ref{fig:Fig2}d) the surface shape is still diamond-like because the parallelogram growth is oriented towards the substrate. These sketches are snapshots of the earliest transformation stages. 
The theoretical descriptions of the martensitic nucleus and its growth stages are supported by the experimental results shown in the following section. 

\subsection{Nucleation and growth process observed \emph{in situ} by SEM}

The samples were heated in-situ into the austenite state and then slowly cooled until the first traces of the martensitic microstructure were visible in the SEM (Fig. \ref{fig:Fig3}). Videos of the entire transformation are presented in the supplementary \cite{Supp1}. Two clearly different nucleus shapes can be distinguished for type X and type Y.

In the first sample, a $5\,\mathrm{\mu m}$ thick Ni-Mn-Ga film that shows mostly \textbf{type X} microstructure (Fig \ref{fig:Fig3}a), an isolated diamond-like nucleus was observed which is about $7\,\mathrm{\mu m}$ long and $1\,\mathrm{\mu m}$ wide. It is separated from the austenite by four straight phase boundaries. It is slightly asymmetric perpendicular to its long axis, which indicates that inside the film it is inclined relative to the substrate. 
\begin{figure}[tb]
	\centering
		\includegraphics[width=0.45\columnwidth]{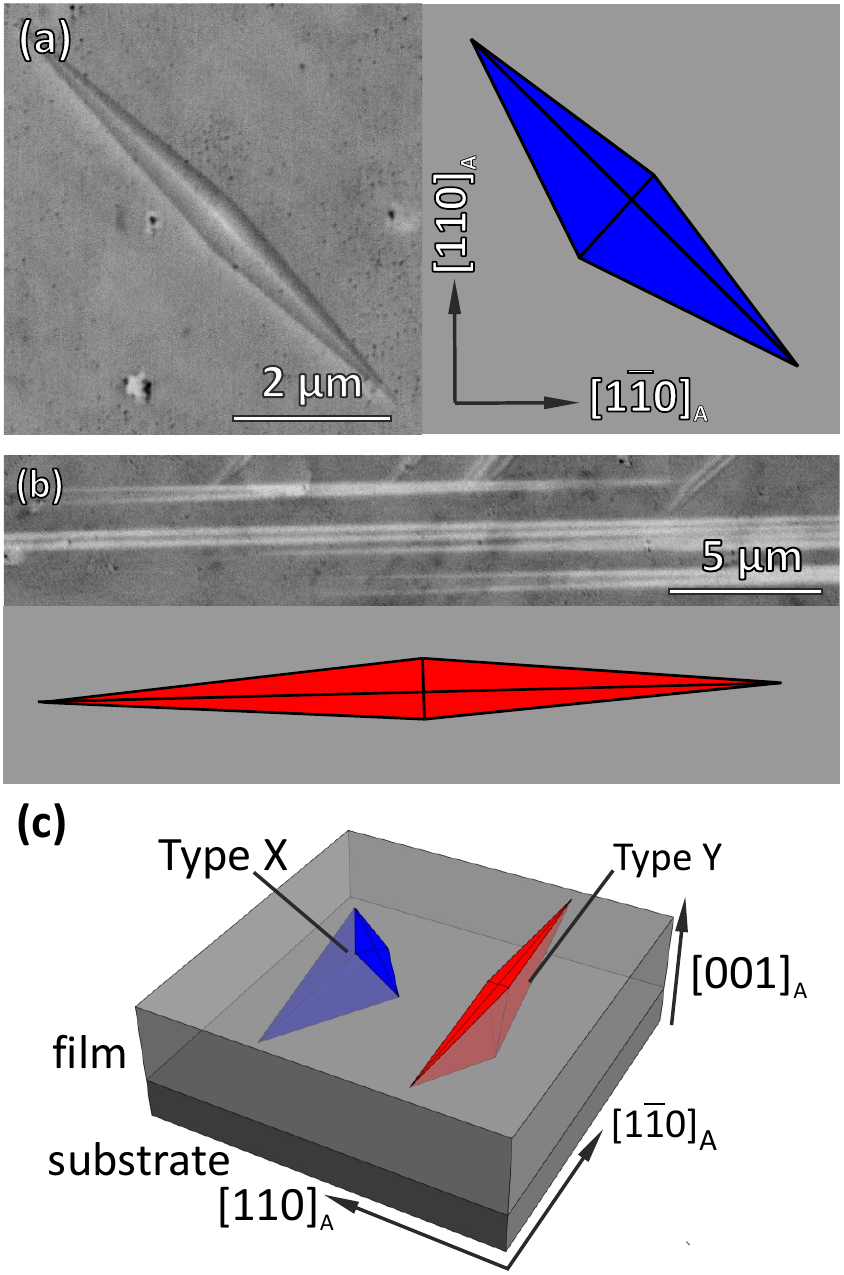}
	\caption{SEM micrographs of the nucleus of type X and type Y martensite. (a) The nucleus of type X was observed by heating the sample into the austenite phase and cooling until first traces of martensite appeared. The object is diamond-like and slightly asymmetric along the long axis. (b) The nucleus of type Y is a needle-like object. (c) Model of the martensitic nucleus relative to the epitaxial film geometry: type X-nuclei are oriented with their long axis along [101], while type Y nuclei are oriented along [110]. The high aspect ratio leads to completely different surface traces as sketched in (a) and (b). }
	\label{fig:Fig3}
\end{figure}
In Fig. \ref{fig:Fig3}b, an early stage of the nucleus of the \textbf{type Y} microstructure from the second sample, a $1.5\,\mathrm{\mu m}$ thick Ni-Mn-Ga film, is presented. The needle-shape object has one tip at the right end, and two tips on the left end.  It is not possible to resolve an inner structure from this SEM image, but it is very likely that these are two nuclei very close together. 

These observations are compared to the theoretical model of the nucleus in Fig. \ref{fig:Fig3}c. The nucleus model from Fig. \ref{fig:Fig2} is sketched relative to the film on the substrate for two orientations. The nuclei are clipped at the film surface to make the surface traces visible. This is reasonable because nuclei may form at the surface which avoids an additional internal interface. 
For type X, the intermediate axis of the diamond is lying in the film plane and the long axis is oriented along $[101]_\mathrm{A}$, which leads to a diamond-like surface feature. For type Y, the long axis of the nucleus is lying in plane along $[1\bar{1}0]_\mathrm{A}$, which leads to a needle-like surface trace because of the very high aspect ratio between the short and the long axis of the diamond.  
At this point, the observations of the nuclei are in good agreement with the proposed three-dimensional nucleus model.

In the following, the transformation path from the nucleation stage to the final microstructure is described.
\begin{figure*}[tb]
	\centering
		\includegraphics[width=\textwidth]{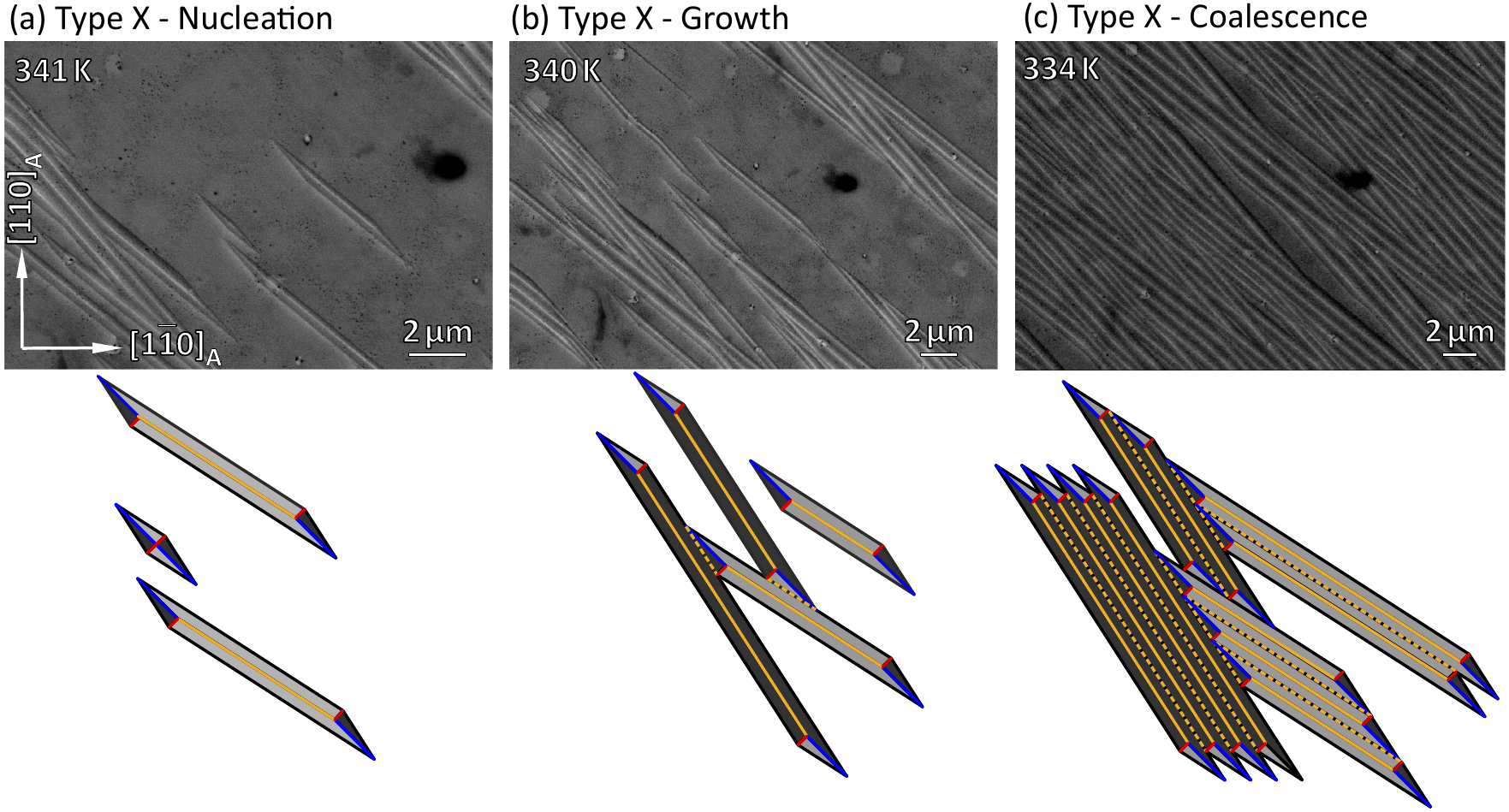}
	\caption{In situ observation of the growth stages of type X martensite and geometrical sketches of the mesoscopic microstructure. The martensite fraction increases from the nucleation stage (a) via the low-density (b) and high-density coalescence step (c). The bottom row sketches the involved twin boundaries using the color code from Fig. \ref{fig:Fig2}.}
	\label{fig:Fig4}
\end{figure*}
The growth of type X is shown in Fig. \ref{fig:Fig4} in the form of scanning electron micrographs at decreasing temperatures in the transformation interval. 
Fig. \ref{fig:Fig4}a was taken at $341\,\mathrm{K}$ during the onset of the transformation. In the middle  of the micrograph, an isolated nucleus is visible, but also a single parallelogram and longer variants. The parallelogram is exactly the growth mode derived from the model in Fig. \ref{fig:Fig2}b and \ref{fig:Fig2}c. The diamond nucleus cannot become very large, because it will touch the substrate at one point. There is no invariant interface to the $(001)_\mathrm{A}$ oriented substrate and thus growth stops in this direction. In order to increase the volume of martensite further, a nucleus can only grow in the film plane via the parallelogram growth mode. In the bottom part of Fig. \ref{fig:Fig4}a, a schematic drawing of the observed microstructure is shown with the same color code for the twin boundaries as in Fig. \ref{fig:Fig2}. The largest twin boundary is the type II as a midrib of the parallelogram. Only near the tips, type I and modulation boundaries are preserved.
In Fig. \ref{fig:Fig4}b, the same spot is shown after cooling by another $1\,\mathrm{K}$. The original diamond nucleus has grown homogenously, which means it has increased its volume while it preserved the same shape. The isolated parallelogram from \ref{fig:Fig4}a is still present but has grown significantly by increasing its length. A high number of new type X parallelograms have appeared. They do not cross each other, since this would not increase the volume transformed, which is the only driving force for the martensitic transformation. During the final stage of coalescence (Fig. \ref{fig:Fig4}c) almost the entire sample is filled with parallelograms. This means that a lot of new parallelograms had to nucleate. They are all aligned parallel, even if there was initially a large distance between them. Thus there must be some kind of long-range interaction between nuclei favoring a parallel alignment. We propose that this is the elastic stress originating from the volume change at the martensitic transformation which cannot be compensated by variant reorientation. A study on this aspect will be published elsewhere. No diamond nuclei are observed in the final stage, so they all transformed into the parallelograms of type X. During coalescence, neighboring parallelograms formed compatible twin boundaries because of the symmetry relation between the meeting variants. These habit planes then vanish and are replaced by compatible twin boundaries. In some cases, the habit plane variants may even be identical and thus during coalescence the interface vanishes. Thus finally, a laminate of mesoscopic twin boundaries forms. The angle between the different type X groups can also be explained by the model, because it originates from the two equivalent growth directions of a given nucleus. Both growth directions are equally allowed, as sketched in Fig. \ref{fig:Fig4}b and c. The experimentally observed angle between them is about 12$\degree$ which is in good agreement with the theoretical angle of $11.3\degree$ calculated from the phenomenological theory of martensite. 

\begin{figure*}[tb]
	\centering
		\includegraphics[width=\textwidth]{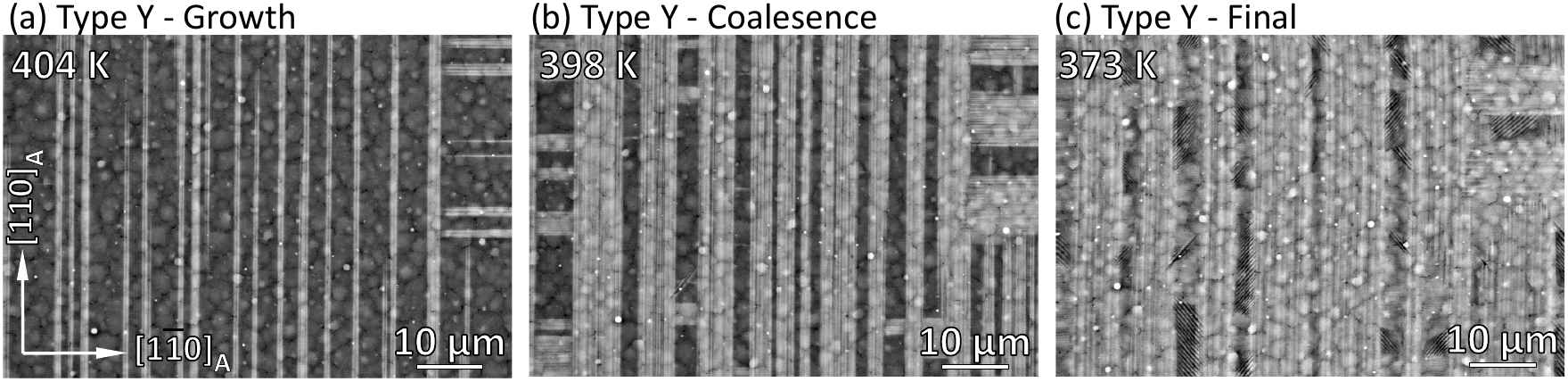}
	\caption{In situ observation of the growth of type Y. (a) Initial microstructure after the onset of the transformation. (b) Parallel groups of needles form and do not intersect each other. In the final stage, type X fills the windows of the martensite.}
	\label{fig:Fig5}
\end{figure*}

The growth of type Y is summarized in Fig. \ref{fig:Fig5}. The type Y nuclei are hardly ever observed in an isolated state, they already form very long needles in the earliest stage of the transformation. Only two major growth directions were observed: Exactly along $[110]_\mathrm{A}$ and exactly along $[1\bar{1}0]_\mathrm{A}$. These needles did not cross each other and neither grew significantly in width. Instead, new needles formed mostly parallel to existing needles. By this process, the austenite fragments into smaller and smaller ``windows''. Interestingly, these windows were filled with type X at the end of the transformation. An example with larger windows compared to Fig. \ref{fig:Fig5} is shown in Fig. \ref{fig:Fig1}a. This supports the idea, that the local stress field around existing martensite favors specific variant orientations.
The reason why no parallelograms were observed in type Y becomes clear when comparing with Fig. \ref{fig:Fig2}d: The parallelogram growth mode also takes place in type Y, but is directed towards the substrate, while the part parallel to the surface remains unchanged. Indeed the cross-section of type Y (Fig. \ref{fig:Fig1}d) shows features of the parallelogram microstructure which resembles the in-plane microstructure of type X. 

\subsection{Macroscopic twin boundaries}
In the previous section, it was described how the nucleus emerges, grows and forms laminates. Now, we will describe what happens when differently oriented laminates of parallelograms meet. 

\begin{figure*}[tb]
	\centering
		\includegraphics[width=\textwidth]{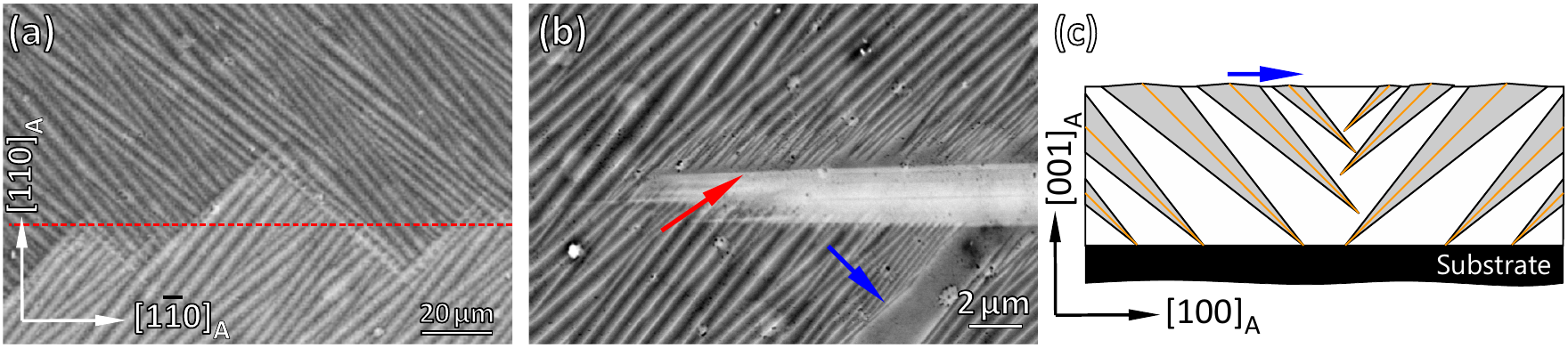}
	\caption{(a) Facetted macroscopic interface between variants of type X martensite at room temperature. The $[1\bar{1}0]_\mathrm{A}$ direction lies in the mirror plane (dashed red). (b) Type X-Y (red arrow) and X-X  (blue arrow) interfaces in the mixed state just above the martensite finish temperature. (c) Model of the microstructure of the X-X-boundary from (b) sketched as a film cross-section below the blue arrow from (b).}
	\label{fig:Fig6}
\end{figure*}

Due to the different equivalent orientations possible for nucleation, a large variety of martensite-martensite interfaces are possible. Usually, growing martensite variants cannot cross each other because this would not increase the volume of transformed martensite. Additionally, not all variants can form twin boundaries with each other. Accordingly, nuclei eventually stop their growth once they meet other nuclei. This is most evident for the horizontal type Y variants in Fig. \ref{fig:Fig5}. These interfaces are called macroscopic twin boundaries, since they connect large martensitic macrovariants and have a width of several micrometers. 
These twin boundaries can be seen in Fig. \ref{fig:Fig1}b, where they connect X and X, Y and Y as well as X and Y laminates. These twin boundaries are a consequence of the kinetic growth process and not just to relieve stresses. Macroscopic twin boundaries originate from different nuclei and, thus, have different habit planes. Accordingly, they are not necessarily compatible to each other and faceting (Fig. \ref{fig:Fig6}a) or branching (Fig. \ref{fig:Fig6}b) occur to minimize interface energy. These twins are obviously not twins on an atomic level, but on a continuum level in accordance with the phenomenological theory of martensite. We propose that some residual austenite persists in this region, which facilitates the reverse transformation. Residual austenite was indeed reported for martensitic Ni-Mn-Ga films, but related to the interface to the substrate without experimental support \cite{Buschbeck2009, Ge_Acta10}.  

Fig. \ref{fig:Fig6}a shows a macroscopic X-X-interface of the $5\,\mathrm{\mu m}$ thick film 1. Each side has formed via the same nucleation and growth process as discussed above. The variants have met in an arrangement that resembles a facetted boundary, which is oriented on average along $[1\bar{1}0]_\mathrm{A}$ and forms the mirror plane (dashed red) between the top and the bottom variants. Hence this represents the third order of twinning. 

Fig. \ref{fig:Fig6}b shows an interface of X and Y martensite also of film 1 at a temperature where the transformation is not completely finished. The parallelograms of X touch the Y-needle with their tips (red arrow), leaving many small disconnected austenitic regions. There, additional martensite can form by new nuclei or a more complex branching process.
The same Fig. \ref{fig:Fig6}b also shows another type of X-X interface (blue arrow). Here, the variants have the same orientation at the surface, but get strongly refined the closer they get to the interface. A sketch of these interfaces is shown in Fig. \ref{fig:Fig6}c. The needles left and right of the macroscopic twin boundary have different growth directions and touch each other within the film. This can explain the large amount of austenite still visible and why the variants become thinner the closer they are located to the interface. Furthermore in an earlier AFM study \cite{Buschbeck2009} we observed the formation of this particular macroscopic twin boundary in-situ. 

\section{Discussion}


To explain the formation of a hierarchical twin-within-twins microstructure we started with a model of the nucleus that is in accordance with the observed early stages of the transformation. We introduced a geometrical model of the martensitic nucleus and its growth mode that was derived from the phenomenological theory of martensite. The diamond nucleus is enclosed by up to eight different phase boundaries. It can form using less variants near a free surface, which is energetically more favorable. It can transform into a parallelogram while keeping the same interface orientations. For thin films this transformation is necessary once a diamond touches the incompatible substrate. The orientation of these interfaces is fixed by the tendency to form stress-free interfaces. In-situ imaging of the martensitic transformation in Ni-Mn-Ga films confirmed the model. The two different observed shapes for the nuclei at the surface of type X and type Y can both be explained taking into account the high aspect ratio of the diamond nucleus. Both, the diamond and the parallelogram, were observed during the initial stages of the martensitic transformation in type X. In the final stage, only the parallelograms were visible. Type Y is dominated by needle-shaped variants that belong to the long axis of the nucleus. 

In this way, a geometrical model using only the basic assumption of the phenomenological theory of martensite - that only certain low-energy interfaces are allowed - can explain most surface features observed in the martensitic state of Ni-Mn-Ga (and isostructural Ni-Mn-X) films. It can explain the orientation of twin variants in both, type Y and type X, the typical angle in type X, the type of twin boundaries observed, and the formation of macroscopic twin boundaries. 

In combination with the model of adaptive martensite \cite{Khach1991a, Kaufmann_PRL10}, our model can explain the observed three levels of twin hierarchy: Nanotwins form directly at the phase boundary and are the only twin bounadries following the classical minimization of twin boundary energy and elastic energy. The mesocopic twin boundaries, which are most decisive for all magnetic shape memory functionality, form as the central planes of symmetry of the martensitic nuclei and are either type I (in a diamond) or type II (in a parallelogram). 
They are required to form a self-accommodated nucleus that is compatible to the surrounding austenite.  As nuclei occur mostly in parallel, resulting in laminates, there must be some kind of elastic interaction, which requires further studies. Finally, macroscopic twin boundaries form between differently oriented nuclei.
With this, the macroscopic length scale of the hierarchical microstructure is essentially governed by nucleation and growth. The system tries to find the most efficient way to transform the material on a local scale. This is in striking difference to the classical picture of Roytburd \cite{Roytburd_93}, who proposed a hierarchical microstructure as a result of global minimization of energy. From our in-situ experiments, we can clearly exclude a global minimization since we did not observe any rearrangement of already transformed material, even though we cooled the material really slow because of the long SEM acquisition times for the in-situ analysis. 

We expect that we can apply this concept to polycrystalline bulk as well: In this case, incompatible grain boundaries take the role of an incompatible substrate and limit the growth of diamonds and parallelograms. This leaves some untransformed regions and we propose that they are transformed during further cooling. Accordingly, the microstructure at macroscopic twin boundaries and incompatible interfaces can be quite complex and will require further studies at higher resolution. 


To conclude, we present in-situ observations of the martensitic transformation in epitaxial Ni-Mn-Ga films that, paired with a simple geometrical model, can explain the formation of most features of the martensitic microstructure including the hierarchy and the formation of macrointerfaces.

\section*{Acknowledgements}
The authors gratefully acknowledge funding by DFG via SPP 1599 and FA 453/8. The work of H.S. and O.H. was supported by Czech Science Foundation (project AdMat No. 14-36566G). 
We thank S.~Schwabe and A.~Diestel for helpful discussions.

\bibliographystyle{ActaMatnew-2}
\bibliography{NiMnGa}

\end{document}